\documentclass[pra]{revtex4-2}

\usepackage{graphicx}
\usepackage{dcolumn}
\usepackage{bm}

\begin{document}
\title{Complete spectral characterization of biphotons by simultaneously determining its frequency sum and difference in a single quantum interferometer}

\author{Baihong Li}
\email{li-baihong@163.com}

\author{Changhua Chen}
\author{Boxin Yuan}
\affiliation{%
Department of Physics, Shaanxi University of Science and Technology, Xi’an 710021, China
}%

\author{Xiangying Hao}%
\author{Rui-Bo Jin}%
 \email{jin@wit.edu.cn}
\affiliation{%
 Hubei Key Laboratory of Optical Information and Pattern Recognition, Wuhan Institute of Technology,
Wuhan 430205, China
}%

\begin{abstract}
We theoretically propose a novel quantum interferometer in which the NOON state interferometer (NOONI) is combined with the Hong-Ou-Mandel interferometer (HOMI). This interferometer combined the advantages of both the NOONI that depends on biphoton frequency sum, and the HOMI that depends on biphoton frequency difference into a single interferometer. It can thus simultaneously obtain the spectral correlation information of biphotons in both frequency sum and difference by taking the Fourier transform from a single time-domain quantum interferogram, which provides a method for complete spectral characterization of an arbitrary two-photon state with exchange symmetry. A direct application of such an interferometer can be found in quantum Fourier-transform spectroscopy where direct spectral measurement is difficult. Furthermore, as it can realize the measurement of time intervals on three scales at the same time, we expect that it can provide a new method in quantum metrology. Finally, we discuss another potential application of such an interferometer in the generation and characterization of high-dimensional and phase-controlled frequency entanglement.

\end{abstract}

\maketitle


\section{\label{sec:1}INTRODUCTION}

Quantum interferometry \cite{Pan}, the heart of various quantum technology applications, can be realized by different quantum interferometer such as the Hong-Ou-Mandel interferometer (HOMI)\cite{PRL1987,RPP2021} and the NOON state interferometer (NOONI)\cite{NOON}. These interferometers have been used to demonstrate various non-classical features of entangled photons such as the violation of Bell’s inequality \cite{PRL1988} and dispersion cancellation \cite{PRA1992,PRL2009,OE2017}. The HOMI has many important applications in quantum information science, e.g., quantum communication \cite{communication}, quantum computing \cite{computer}, quantum imaging \cite{Imaging1,Imaging} and quantum metrology \cite{metrology1,metrology2,attosecond}. The NOONI has been widely used in quantum lithography \cite{NOON,NOON1}, quantum high-precision measurement \cite{NOON2,NOON3,NOON4}, quantum microscopy \cite{microscopy1,microscopy2,microscopy3}, error correction \cite{error} and so on.

It is well known that using the interferometric spectrometer technology established by the Wiener–Khinchin theorem in classical optics, it is possible to extract the spectral information of light by making a Fourier transform on its time-domain interferograms obtained from Mach-Zehnder or Michelson interferometers. Jin et.al \cite{JIN2018,PRApplied2022} extended the Wiener-Khinchin theorem to a quantum version where the biphoton spectral information of frequency difference and frequency sum between signal and idler photons can be extracted by applying a Fourier transform on the time-domain patterns of the HOMI and NOONI, respectively. This can be considered as a kind of quantum Fourier-transform spectroscopy. However, the HOMI only depends on frequency difference, and the NOONI only depends on frequency sum, making these two interferometers access only one-dimension (1D) information  in frequency difference or frequency sum, which is mismatched with two-photon states in two dimensions \cite{JIN2018,PRA2002}. To solve this problem, Abouraddy et.al \cite{PRA2013} theoretically  proposed a linear two-photon interferometer containing two independent delays but it is relatively hard to realize in an experiment. Alternatively, a 2D joint spectral intensity (JSI) associated with the frequency sum and difference and its Fourier transform, joint temporal intensity (JTI), associated with the time difference and sum can be obtained directly in an experiment but it is especially challenging and impractical especially for the broadband spectrum \cite{PRL2018,PRApplied2018}. A comparative study of various different techniques on spectral characterization of biphotons, as well as their relative advantages and disadvantages can be found in \cite{JMO2018}.

In this paper, we propose a new type of quantum interferometer in which the NOONI is combined with the HOMI. This interferometer depends on both frequency sum and frequency difference and combines the advantages of both the NOONI and the HOMI into a single interferometer. The temproal interference patterns associated with biphoton frequency sum and difference can be shown in different parts of an interferogram. It can thus obtain simultaneously the spectral correlation information of biphotons both in frequency sum and frequency by taking the Fourier transform of the time-domain quantum interferograms obtained from such an interferometer. This may be especially useful for quantum Fourier-transform spectroscopy where direct spectral measurement is difficult. The specific interferograms of the interferometer depend on the exchange symmetry of biphoton and the ratio between the width of frequency difference and frequency sum, which correspond to different types of frequency entangled resources, such as frequency correlated, uncorrelated or anti-correlated. Therefore, the proposed interferometer can be used for complete spectral characterization of an arbitrary two-photon state with exchange symmetry. Moreover, as it can realize the measurement of time intervals on three scales at the same time, we expect that it can provide a new method in quantum metrology. Finally, we discuss another potential application of such an interferometer in the generation and characterization of high-dimensional and phase-controlled frequency entanglement.

The rest of the paper is organized as follows. In section \ref{sec:2}, we describe the setup of the proposed interferometer and discuss several interferometric results based on its coincidence count rates derived theoretically from frequency domain. In section \ref{sec:3}, we give some typical interferograms of the interferometer for different types of frequency entangled resources and make a characteristic analysis of the interferometer. In section \ref{sec:4}, we compare our results with that of the NOONI and HOMI, and discuss potential applications of such interferometer in quantum Fourier-transform spectroscopy, quantum metrology and generation and characterization of high-dimensional and phase-controlled frequency entanglement. Section \ref{sec:conclude} summarizes the results and concludes the paper.

\section{\label{sec:2}Theory of the combination interferometer}

\begin{figure}[th]
\begin{picture}(400,140)
\put(0,0){\makebox(405,120){
\scalebox{0.9}[0.9]{
\includegraphics{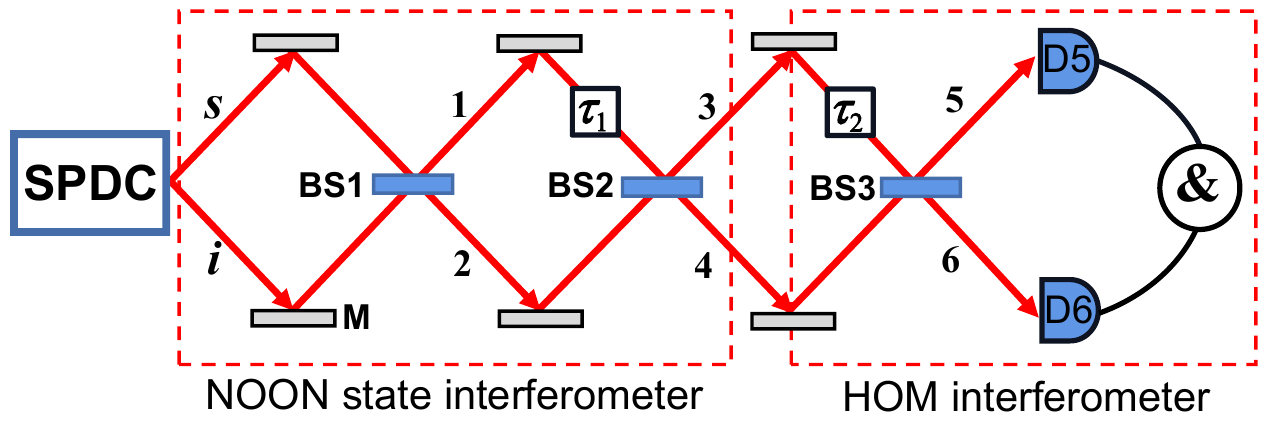}
}}}
\end{picture}
\caption{\label{Fig1}
Schematic diagram of the proposed quantum interferometer. It can be considered as a combination of the NOON state interferometer (the left part) and the HOM interferometer (the right part). M:Mirror, BS: beamspliter, D: detector, \boldsymbol{\&} : coincidence count.}
\end{figure}

In this part, we propose a novel quantum interferometer with the setup shown in Fig. \ref{Fig1}. To understand better such an interferometer, it can be considered as a combination of the NOONI (the left part) and the HOMI (the right part). The biphotons are generated by the spontaneous parametric down-conversion (SPDC) or spontaneous four-wave-mixing process. As derived in Appendix A, the coincidence count rates between two detectors (D5 and D6) as functions of time delay $\tau_1$ and $\tau_2$ for the combination interferometer can be expressed as
\begin{equation}
\label{R}
R(\tau_1,\tau_2)= \frac{1}{64}\int_{0}^{\infty}\int_{0}^{\infty}d\omega_s d\omega_i r(\omega_s,\omega_i,\tau).
\end{equation}
where $r$ is the coincidence probability density, which reads
\begin{eqnarray}
\label{r}
r(\omega_s,\omega_i,\tau_1,\tau_2)=&&|f(\omega_s,\omega_i)(e^{-i\omega_s (\tau_1+\tau_2)}+e^{-i\omega_s \tau_1}+e^{-i\omega_s \tau_2}-1)(e^{-i\omega_i (\tau_1+\tau_2)}-e^{-i\omega_i \tau_1}-e^{-i\omega_i \tau_2}-1)\nonumber\\
&&+f(\omega_i,\omega_s)(e^{-i\omega_i (\tau_1+\tau_2)}+e^{-i\omega_i \tau_2}-e^{-i\omega_i \tau_1}+1)(e^{-i\omega_s (\tau_1+\tau_2)}-e^{-i\omega_s \tau_2}+e^{-i\omega_s \tau_1}+1)|^2.
\end{eqnarray}
where $f(\omega_s,\omega_i)$ is the joint spectral amplitude (JSA) of the signal and idler photons. The specific expression of $r$ depends on the symmetry of the JSA.  For simplicity, we assume that the JSA is symmetric, i.e., $f(\omega_s,\omega_i)=f(\omega_i,\omega_s)$ in our discussions below. Assuming that $\omega_s=\omega_p/2+\Omega_s$ and $\omega_i=\omega_p/2+\Omega_i$, where $\Omega_{s,i}$ is the frequency detuning between the signal (idler)photon and the half of pump center frequency $\omega_p/2$, then Eq.(\ref{r}) can be expressed as the form of frequency sum $\Omega_{+}=\Omega_s+\Omega_i$ and frequency difference $\Omega_-=\Omega_s-\Omega_i$,
\begin{eqnarray}
\label{r-pulse}
r(\Omega_+,\Omega_-,\tau_1,\tau_2)=32|f(\Omega_+,\Omega_-)|^2 \{1- \frac{1}{2}\cos[(\omega_p+\Omega_+) \tau_1]\cos(\Omega_-\tau_2)- \frac{1}{2}\cos[(\omega_p+\Omega_+)\tau_2]\nonumber\\
-\frac{1}{2}\cos(\Omega_-\tau_2)+\frac{1}{4}\cos[(\omega_p+\Omega_+)(\tau_2+\tau_1)]+\frac{1}{4}\cos[(\omega_p+\Omega_+)(\tau_2-\tau_1)]\}.
\end{eqnarray}
In general, the JSA can not be factorized as a product of $f(\omega_s)$ and $f(\omega_i)$. However, $f(\Omega_+,\Omega_-)$ can be factorized as a product of $f(\Omega_+)$ and $f(\Omega_-)$ in terms of collective coordinate $\Omega_+$ and $\Omega_-$, i.e., $f(\Omega_+,\Omega_-)=f(\Omega_+)f(\Omega_-)$\cite{PRApplied2018}. Eq.(\ref{r-pulse}) can then be integrated independently with respect to $\Omega_+$ and $\Omega_-$.  If we define $F(\Omega_{\pm})=|f(\Omega_{\pm})|^2$, its Fourier transform would be
\begin{equation}
\label{F-T}
G_{\pm}(\tau)=\frac{1}{\sqrt {2\pi}}\int_{-\infty}^{\infty}F(\Omega_{\pm}) e^{i\Omega_{\pm} \tau}d\Omega_{\pm}.
\end{equation}
Integrating Eq.(\ref{r-pulse}) over the entire frequency range, it is found that $R(\tau_1,\tau_2)$ can be expressed as the functions of $G(\tau)$. We can thus obtain the normalized coincidence count rates,
\begin{equation}
\label{Rn}
R_{N}(\tau_1,\tau_2)=1-\frac{1}{2}g_+(\tau_1)g_+(\tau_2)-\frac{1}{2}g_+(\tau_2)-\frac{1}{2}g_-(\tau_2)+ \frac{1}{4}g_+(\tau_2+\tau_1)+ \frac{1}{4}g_+(\tau_2-\tau_1).
\end{equation}
where $g_{\pm}(\tau)=Re[G_{\pm}(\tau)/G_{\pm}(0)]$. Eq.(\ref{Rn}) is the central equation of the present manuscript. It can be seen from Eq.(\ref{r-pulse}) and Eq.(\ref{Rn}) that the result of the combination interferometer is determined by both frequency sum and frequency difference, which is quite different from the NOON state interference determined only by the frequency sum and the standard HOM interference determined only by the frequency difference for the symmetric JSA \cite{JIN2018,PRA2002} (Also see Appendix C). In other words, Eq.(\ref{Rn}) contains complete spectral information of biphotons associated with both frequency sum and difference, which can be obtained by making a Fourier transform of a single time-domain quantum interferogram obtained from the combination interferometer. Additionally, it can be found that if $\tau_1=0$, Eq.(\ref{Rn}) will deduce to that of a standard HOM interference \cite{JIN2018}. On the other hand, if $\tau_2=0$, the coincidence count rates will always be equal to zero.

As an example, we take the symmetric JSA as the products of two Gaussian functions, i.e., $f(\Omega_+,\Omega_-)=\exp(-\Omega_+^2/4\sigma_+^2)\exp(-\Omega_-^2/4\sigma_-^2)$, where $\sigma_{\pm}$ denote the linewidth of two functions determined by the linewidth of pump pulse and the phase-matching condition, respectively. Eq.(\ref{Rn}) now becomes
\begin{eqnarray}
\label{RN}
R_{N}(\tau_1,\tau_2)=1-\frac{1}{2}\cos(\omega_p \tau_1)e^{-\sigma_+^2\tau_1^2/2}e^{-\sigma_-^2\tau_2^2/2}-\frac{1}{2}\cos(\omega_p \tau_2)e^{-\sigma_+^2\tau_2^2/2}-\frac{1}{2}e^{-\sigma_-^2\tau_2^2/2}\nonumber\\
+\frac{1}{4}\cos[\omega_p (\tau_2+\tau_1)]e^{-\sigma_+^2(\tau_2+\tau_1)^2/2}+\frac{1}{4}\cos[\omega_p (\tau_2-\tau_1)]e^{-\sigma_+^2(\tau_2-\tau_1)^2/2}.
\end{eqnarray}
Eq.(\ref{RN}) involves the four time scales, i.e., $\tau_1,\tau_2$ and the inverse linewidths $1/\sigma_+, 1/\sigma_-$. If we set $\tau_1$ to be a fixed value, then Eq.(\ref{RN}) becomes only as a funtion of $\tau_2$. In this case, the last two terms in Eq.(\ref{RN}) correspond to two identical interferograms involving a cosine oscillation with a period of $\omega_p$ centered at $\pm\tau_1$. If $\tau_1<1/\sigma_+$, these two interferograms will gradually overlap and become fully indistinguishable as $\tau_1$ decreases to be zero. Thus, if one would like to distinguish these two interferograms, $\tau_1$ must be much larger than the inverse linewidth $1/\sigma_+$. If so, the second term in Eq.(\ref{RN}) will tend to be zero. Eq.(\ref{RN}) can then be simplified as
\begin{eqnarray}
\label{RN1}
R_{N}(\tau_1,\tau_2)=&&1-\frac{1}{2}\cos(\omega_p \tau_2)e^{-\sigma_+^2\tau_2^2/2}-\frac{1}{2}e^{-\sigma_-^2\tau_2^2/2}+\frac{1}{4}\cos[\omega_p (\tau_2+\tau_1)]e^{-\sigma_+^2(\tau_2+\tau_1)^2/2}\nonumber\\
&&+\frac{1}{4}\cos[\omega_p (\tau_2-\tau_1)]e^{-\sigma_+^2(\tau_2-\tau_1)^2/2}.
\end{eqnarray}
For a fixed value of $\tau_1$, the second term in Eq.(\ref{RN1}) represents a cosine oscillation centered at $\tau_2=0$ with a temporal width of the envelope that is inversely proportional to the linewidth $\sigma_+$ and with an interference visibility of 1/2. The period of the oscillation is determined by the pump center frequency $\omega_p$. The third term in Eq.(\ref{RN1}) corresponds to a standard HOM dip centered at $\tau_2=0$ with a temporal width that is inversely proportional to the linewidth $\sigma_-$ and with an interference visibility of 1/2. The last two terms in Eq.(\ref{RN1}) correspond to two identical oscillations around $\pm\tau_1$ both with a temporal width of the envelope that is inversely proportional to the linewidth $\sigma_+$ and with an interference visibility of 1/4.

If $\tau_2$ is a non-zero constant and $\tau_2>>1/\sigma_+$, Eq.(\ref{RN1}) can be further simplified to
\begin{eqnarray}
\label{RN2}
R_{N}(\tau_1,\tau_2)=1+\frac{1}{4}\cos[\omega_p (\tau_1+\tau_2)]e^{-\sigma_+^2(\tau_2+\tau_1)^2/2}+\frac{1}{4}\cos[\omega_p (\tau_1-\tau_2)]e^{-\sigma_+^2(\tau_2-\tau_1)^2/2}.
\end{eqnarray}
In this case, the coincidence count rate is only the function of $\tau_1$, and the interferogram will only contain two-side oscillations around $\pm\tau_2$ with a temporal width of the envelope that is inversely proportional to the linewidth $\sigma_+$.

\section{\label{sec:3}Characteristic analysis of the combination interferometer}

\begin{figure}[th]
\begin{picture}(380,150)
\put(0,0){\makebox(370,140){
\scalebox{0.5}[0.5]{
\includegraphics{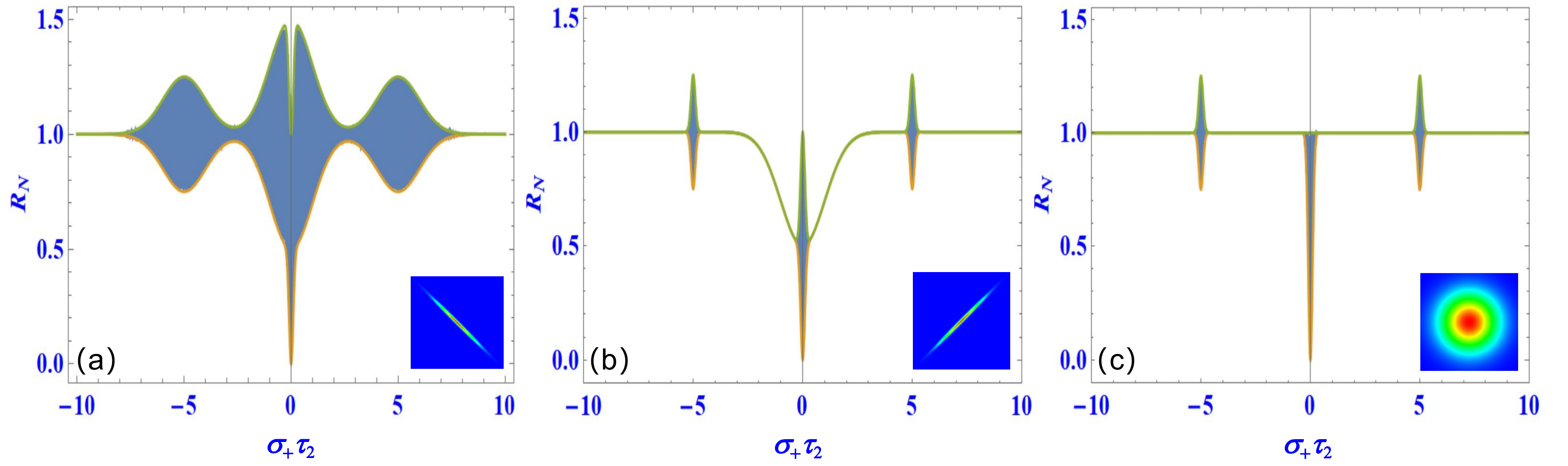}
}}}
\end{picture}
\caption{\label{Fig2}
Typical interferograms of the combination interferometer as a function of $\sigma_+\tau_2$ at $\sigma_+\tau_1=5$ for (a) frequency anti-correlated ($\sigma_+/\sigma_-=0.1$), (b) frequency correlated ($\sigma_+/\sigma_-=10$) and (c) frequency uncorrelated ($\sigma_+/\sigma_-=1$) resources. The temproal interference patterns associated with biphoton frequency sum and difference can be shown in different parts of an interferogram for all type of frequency entangled resources. The green and orange curves denote the upper and lower envelopes of the interferograms, respectively. The time delays are in unit of the
inverse of $\sigma_+$. The insets are the corresponding JSIs with the same  amplitude of horizontal and vertical coordinates.}
\end{figure}

To understand better the characteristic of the combination interferometer, we give some typical interferograms for different types of frequency correlation based on Eq.(\ref{RN1}). Figure \ref{Fig2} shows the typical interferograms for the combination interferometer as a function of $\sigma_+\tau_2$ at $\sigma_+\tau_1=5$ for frequency anti-correlated (a), correlated (b) and uncorrelated (c) resources. We can see that the temproal interference patterns associated with biphoton frequency sum and difference can be shown in different parts of an interferogram for all type of frequency entangled resources. It can thus obtain simultaneously the spectral correlation information of biphotons both in frequency sum and frequency by taking the Fourier transform of the upper (green) or lower (orange) envelopes of the interferograms in Fig.\ref{Fig2}. Additionally, the interferograms in Fig.\ref{Fig2} contain the information of three time scales, i.e., the temporal width of the envelope of two-side interferograms determined by the inverse linewidth $1/\sigma_+$, the temporal width of the middle dip determined by the inverse linewidth $1/\sigma_-$, and the time interval between two-side interferograms determined by $\pm\tau_1$. It can thus realize the measurement of time intervals on three scales at the same time in a single experiment, which might be useful in quantum metrology. For frequency anti-correlated resource (Figure \ref{Fig2}(a)), the width of frequency difference is much larger than the width of frequency sum, as a result, there is a narrow dip around $\sigma_+\tau_2=0$ and a wider envelope of two-side interferograms around $\sigma_+\tau_1=\pm5$ due to their inverse dependence with respect to the linewidth of frequency difference and sum, respectively. However, for frequency correlated resource (Figure \ref{Fig2}(b)), the width of frequency sum is much larger than the width of frequency difference, resulting in a wider dip around $\sigma_+\tau_2=0$ and a narrow envelope of two-side interferograms around $\sigma_+\tau_1=\pm5$.  For frequency uncorrelated resource (Figure \ref{Fig2}(c)), the width of frequency sum is identical to the width of frequency difference, resulting in the same width of the middle and the two-side interferograms. Therefore, one can distinguish different types of frequency correlation only by observing a single time-domain quantum interferogram obtained from the combination interferometer. Conversely, one can also characterize different types of frequency correlation by using only a single time-domain quantum interferogram.

\begin{figure}[th]
\begin{picture}(400,180)
\put(0,0){\makebox(370,180){
\scalebox{0.45}[0.45]{
\includegraphics{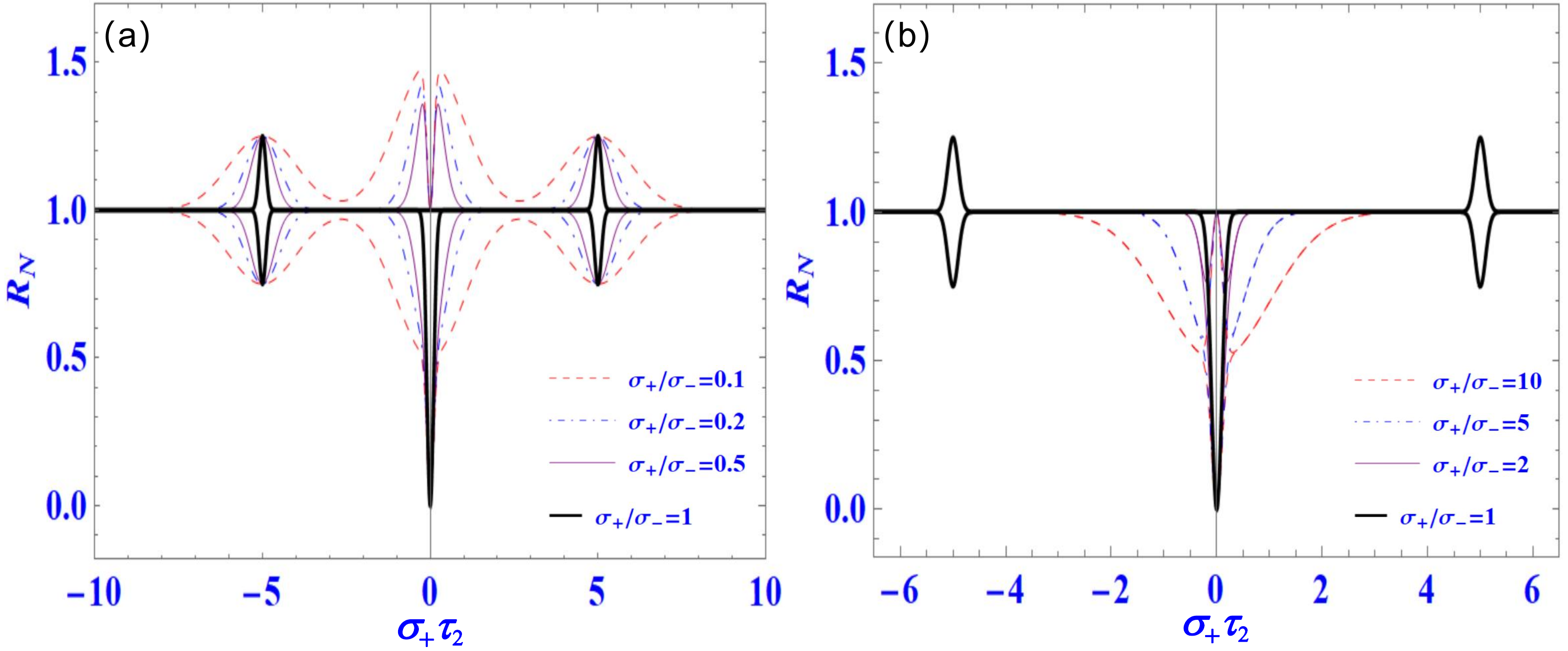}
}}}
\end{picture}
\vspace{0.1cm}
\caption{\label{Fig3}
The envelopes of the interferograms of the combination interferometer as a function of $\sigma_+\tau_2$ at $\sigma_+\tau_1=5$ for (a) frequency anti-correlated and (b) frequency correlated resources at different ratios $\sigma_+/\sigma_-$. The bold black lines correspond to the results of the frequency uncorrelated resources as a contrast.}
\end{figure}
In order to explore the effect of different degree of frequency correlation on the interferogram, we plot the envelopes of the interferograms as a function of $\sigma_+\tau_2$ at $\sigma_+\tau_1=5$ for frequency anti-correlated  and frequency correlated resources at different ratios $\sigma_+/\sigma_-$, as shown in Fig.\ref{Fig3}. For frequency anti-correlated resource (Figure \ref{Fig3}(a)), the width of the two-side envelopes and the central envelope of $R_N$ above 1/2 decreases with the increase of ratios $\sigma_+/\sigma_-$ (equivalent increase of $\sigma_+$), because their widths are only determined by biphoton frequency sum. The central envelope of $R_N$ below 1/2 remains unchanged due to the same width of biphoton frequency difference. The widths tends to be identical as the ratios increase to be one. For frequency correlated resource (Figure \ref{Fig3}(b)), the width of the two-side envelopes remains unchanged due to the fixed value of $\sigma_-$. The central envelope of $R_N$ above 1/2 increases with the increase of ratios $\sigma_+/\sigma_-$ (equivalent decrease of $\sigma_-$), because their widths are only determined by biphoton frequency sum.

\section{\label{sec:4}discussion}
Based on the combination interferometer, we can build a bridge between a 2D spectral correlation information of biphotons containing frequency sum and difference and 1D time-domain interference patterns in a single quantum interferometer.  By directly taking Fourier transform of the envelope of an interferogram obtained from the combination interferometer, one can obtain the complete spectral information of biphotons both in frequency sum and difference, i.e., the distributions of $|f(\Omega_+)|^2$ and $|f(\Omega_-)|^2$. Finally, the JSI can be reconstructed by their product. A direct application of such an interferometer can be found in the field of quantum Fourier-transform spectroscopy \cite{PRApplied2021,PRApplied2022}, especially for those spectra whose spectral range is not easy to obtain by the conventional spectrograph.

It should be noted that it can be seen from Eq.(\ref{F-T}) and Eq.(\ref{Rn}) that, the results of the time-domain interference of the combination interferometer are only determined by the intensity (not the amplitude) of the biphoton field. This holds for the HOMI and the NOONI, too. It means that these interferometers are not sensitive to the phase of the biphoton field, making it more convenient to measure in an experiment. However, the JSA is sensitive to the phase and is difficult to measure experimentally. If one would like to completely characterize the two-photon state, it is necessary to measure not only the amplitude but also the phase of biphoton state, as shown in \cite{PRA2019,Optica2020}.

From the analysis above, we can see that the results of the combination interferometer also depend on the symmetry of the JSA. If the JSA is symmetric, i.e., $f(\omega_s,\omega_i)=f(\omega_i,\omega_s)$, it will result in the bunching effect in the HOMI, a feature associated with bosonic statistics. If the JSA is antisymmetric, i.e., $f(\omega_s,\omega_i)=-f(\omega_i,\omega_s)$, however, this will lead to the antibunching effect, a feature associated with fermionic statistics. These effects have been demonstrated in spatial degree of freedom (DOF)\cite{Walborn03}, frequency DOF \cite{Optica20} and both DOFs of polarization and orbital angular momentum \cite{Wang20}. Also, the influence of the exchange symmetry of the biphoton on the coincidence measurement of both the HOMI and the NOONI have been studied recently in \cite{Fabre22}. For the HOMI, coincidence counts always depend on frequency difference of biphotons whatever the JSA is symmetric, antisymmetric or anyonic, and a continuous deformation of coincidence counts happens from a dip obtained with symmetric JSA to a peak obtained with antisymmetric JSA. However, this is not the case for the NOONI where it depends on biphoton frequency sum for the symmetric JSA but frequency difference for the antisymmetric JSA \cite{Fabre22}. In our case, however, the combination interferometer will always depend on both frequency difference and sum whatever the JSA is symmetric, antisymmetric or anyonic. Similarly, the change of the exchange symmetry of the biphoton from symmetric to antisymmetric will flip the interferograms in Fig.\ref{Fig2} along $y>0$ axis at the baseline of $R_N=1$. Therefore, the combination interferometer can be extended to completely characterize spectral feature of biphotons both in frequency difference and sum for any symmetry of the JSA.

Now let us consider another potential application of such an interferometer in generation and characterization of high-dimensional frequency entanglement with the help of the spectrally resolved technology \cite{SR-PRA2015,SR-OE2015}. In the spectrally resolved HOM interference, the JSI can be modulated along the axis of frequency difference for the frequency anti-correlated entangled resource, which has been proved to be very useful for generating high-dimensional entanglement in frequency-bin qudits \cite{jin2016,npj2021,OLT2023}. On the other hand, in spectrally resolved NOON state interference, it has been demonstrated that the JSI can be modulated by the time delay along the directions of the frequency sum between the signal and idler photons\cite{3-photons2019,SR-NOON2021}. Therefore, it is also possible to generate high-dimensional frequency entanglement using the left part of the combination interferometer in Fig.\ref{Fig1} with the help of the spectrally resolved technology. This can be realized by modulating the JSI along the axis of the frequency sum for frequency correlated resource via adjusting the time delay $\tau_1$. Since the NOONI depends on biphoton frequency sum, which is thus phase-dependent, the generated frequency entanglement from this way is phase-controlled. In principle, an arbitrary-dimensional discrete frequency entanglement can be prepared by adjusting the time delay $\tau_1$ and the dimensionality of frequency entanglement increase as the increase of time delay $\tau_1$. Meanwhile, one can use the right part(HOMI) of the combination interferometer in Fig.\ref{Fig1} to characterize directly the generated frequency entanglement, according to the time interval 2$\tau_1$ between two-side temporal interferograms. It means that the larger the time interval 2$\tau_1$, the higher the dimensionality of frequency entanglement.

\section{\label{sec:conclude}CONCLUSIONS}
We have proposed a novel combination interferometer that always depends on both biphoton frequency sum and difference whatever the exchange symmetry of biphotons. With this interferometer, it is possible to simultaneously obtain the spectral correlation information of biphotons in both frequency sum and difference by taking the Fourier transform of a single time-domain quantum interferogram, which provides a method for complete spectral characterization of an arbitrary two-photon state with exchange symmetry and might be useful in quantum Fourier-transform spectroscopy where direct spectral measurement is difficult. The typical interferograms for different types of frequency correlation have been presented to show the characteristic of the combination interferometer. Furthermore, as it can realize the measurement of time intervals on three scales at the same time, we expect that it can provide a new method in quantum metrology. Finally, as a potential application, we have shown that it is also possible to generate high-dimensional and phase-controlled frequency entanglement in such an interferometer with spectrally resolved technology by adjusting the time delay and characterize it directly with the two-side oscillations appeared in temporal interferogram.

\begin{acknowledgments}
This work has been supported by National Natural Science Foundation of China (12074309, 12074299), and the Youth Innovation Team of Shaanxi Universities.
\end{acknowledgments}

\vspace{0.5cm}

\appendix

\section{The coincidence count rates for the combination interferometer derived from the frequency domain}
In this Section, we deduce the equations for the combination interferometer in Fig. \ref{Fig1} from the frequency domain. The two-photon state from a SPDC process can be described as
\begin{equation}
|\Psi\rangle =\int\int d\omega_s d\omega_i f(\omega_s, \omega_i)\hat{a}_s^\dag(\omega_s)\hat{a}_i^\dag(\omega_i)|0\rangle
\end{equation}
where $\omega$ is the angular frequency, and $\hat{a}_{s,i}^\dag$ is the creation operator and the subscripts $s$ and $i$ denote the signal and idler photons from SPDC, respectively. $|0\rangle$ stands for a vacuum state. $f(\omega_s, \omega_i)$ is the JSA of the signal and idler photons.

The detection field operators of detector 5 (D5) and detector 6 (D6) are
\begin{equation}
\hat{E}_5^{(+)}(t_5)= \frac{1}{\sqrt{2\pi}}\int_{0}^{\infty}d\omega_5\hat{a}_5(\omega_5)e^{-i\omega_5t_5}
\end{equation}
\begin{equation}
\hat{E}_6^{(+)}(t_6)= \frac{1}{\sqrt{2\pi}}\int_{0}^{\infty}d\omega_6\hat{a}_6(\omega_6)e^{-i\omega_6t_6}
\end{equation}
where the subscripts 5 and 6 denote the photons detected by D5 and D6, respectively.  The transformation rule of the 50/50 beamsplitter (BS3) is
\begin{equation}
\hat{a}_5(\omega_5)=[\hat{a}_3(\omega_5)e^{-i\omega_5\tau_2}+\hat{a}_4(\omega_5)]/\sqrt{2}
\end{equation}
\begin{equation}
\hat{a}_6(\omega_6)=[\hat{a}_3(\omega_6)e^{-i\omega_6\tau_2}-\hat{a}_4(\omega_6)]/\sqrt{2}
\end{equation}
The transformation rule of the 50/50 BS2 is
\begin{equation}
\hat{a}_3(\omega_5)=[\hat{a}_1(\omega_5)e^{-i\omega_5\tau_1}+\hat{a}_2(\omega_5)]/\sqrt{2}
\end{equation}
\begin{equation}
\hat{a}_4(\omega_6)=[\hat{a}_1(\omega_6)e^{-i\omega_6\tau_1}-\hat{a}_2(\omega_6)]/\sqrt{2}
\end{equation}
The transformation rule of the 50/50 BS1 is
\begin{equation}
\hat{a}_1(\omega_5)=[\hat{a}_s(\omega_5)+\hat{a}_i(\omega_5)]/\sqrt{2},\hat{a}_2(\omega_5)=[\hat{a}_s(\omega_5)-\hat{a}_i(\omega_5)]/\sqrt{2}.
\end{equation}
\begin{equation}
\hat{a}_1(\omega_6)=[\hat{a}_s(\omega_6)+\hat{a}_i(\omega_6)]/\sqrt{2},\hat{a}_2(\omega_6)=[\hat{a}_s(\omega_6)-\hat{a}_i(\omega_6)]/\sqrt{2}.
\end{equation}
So, we have
\begin{equation}
\hat{a}_3(\omega_5)=[(\hat{a}_s(\omega_5)+\hat{a}_i(\omega_5))e^{-i\omega_5\tau_1}+(\hat{a}_s(\omega_5)-\hat{a}_i(\omega_5))]/2,
\end{equation}
\begin{equation}
\hat{a}_4(\omega_5)=[(\hat{a}_s(\omega_5)+\hat{a}_i(\omega_5))e^{-i\omega_5\tau_1}-(\hat{a}_s(\omega_5)-\hat{a}_i(\omega_5))]/2,
\end{equation}
\begin{equation}
\hat{a}_3(\omega_6)=[(\hat{a}_s(\omega_6)+\hat{a}_i(\omega_6))e^{-i\omega_6\tau_1}+(\hat{a}_s(\omega_6)-\hat{a}_i(\omega_6))]/2,
\end{equation}
\begin{equation}
\hat{a}_4(\omega_6)=[(\hat{a}_s(\omega_6)+\hat{a}_i(\omega_6))e^{-i\omega_6\tau_1}-(\hat{a}_s(\omega_6)-\hat{a}_i(\omega_6))]/2.
\end{equation}
Substituting Eq.(A10)-Eq.(A13) into Eq.(A4)-Eq.(A5), we have
\begin{eqnarray}
\hat{a}_5(\omega_5)=[\hat{a}_s(\omega_5)(e^{-i\omega_5 (\tau_1+\tau_2)}+e^{-i\omega_5 \tau_1}+e^{-i\omega_5 \tau_2}-1)+\hat{a}_i(\omega_5)(e^{-i\omega_5 (\tau_1+\tau_2)}+e^{-i\omega_5 \tau_1}-e^{-i\omega_5 \tau_2}+1)]/2\sqrt{2}
\end{eqnarray}
\begin{eqnarray}
\hat{a}_6(\omega_6)=[\hat{a}_s(\omega_6)(e^{-i\omega_6 (\tau_1+\tau_2)}+e^{-i\omega_6 \tau_2}-e^{-i\omega_6 \tau_1}+1)+\hat{a}_i(\omega_6)(e^{-i\omega_6 (\tau_1+\tau_2)}-e^{-i\omega_6 \tau_2}-e^{-i\omega_6 \tau_1}-1)]/2\sqrt{2}
\end{eqnarray}
The coincidence count rates between two detectors as functions of delay time $\tau_1,\tau_2$ can be expressed as
\begin{equation}
R(\tau_1,\tau_2)= \int \int d t_5 d t_6 \langle \Psi |\hat{E}_5^{(-)}\hat{E}_6^{(-)} \hat{E}_6^{(+)}\hat{E}_5^{(+)}|\Psi \rangle=\int \int d t_5 d t_6 |\langle 0| \hat{E}_6^{(+)}\hat{E}_5^{(+)}|\Psi \rangle|^2
\end{equation}
Consider $ \hat{E}_6^{(+)}\hat{E}_5^{(+)}|\Psi \rangle$, only 2 out of 4 terms exist. The first term is
\begin{eqnarray}
&&\frac{1}{16\pi}\int\int d\omega_5d\omega_6\hat{a}_s(\omega_5)\hat{a}_i(\omega_6)e^{-i\omega_5 t_5}e^{-i\omega_6 t_6}(e^{-i\omega_5 (\tau_1+\tau_2)}+e^{-i\omega_5 \tau_1}+e^{-i\omega_5 \tau_2}-1)\nonumber\\
&&\times (e^{-i\omega_6 (\tau_1+\tau_2)}-e^{-i\omega_6 \tau_2}-e^{-i\omega_6 \tau_1}-1)\int \int d\omega_s d\omega_i f(\omega_s, \omega_i)\hat{a}_s^\dag(\omega_s)\hat{a}_i^\dag(\omega_i)|0\rangle \nonumber\\
&&= \frac{1}{16\pi} \int \int d\omega_5 d\omega_6 e^{-i\omega_5 t_5}e^{-i\omega_6 t_6} f(\omega_5, \omega_6)(e^{-i\omega_5 (\tau_1+\tau_2)}+e^{-i\omega_5 \tau_1}+e^{-i\omega_5 \tau_2}-1)
\nonumber\\
&&\times (e^{-i\omega_6 (\tau_1+\tau_2)}-e^{-i\omega_6 \tau_2}-e^{-i\omega_6 \tau_1}-1)|0\rangle \nonumber\\
\end{eqnarray}
In this calculation, the relationship of $\hat{a}_5(\omega_5)\hat{a}_s^\dag(\omega_s)=\delta(\omega_5-\omega_s),\hat{a}_i(\omega_6)\hat{a}_i^\dag(\omega_i)=\delta(\omega_6-\omega_i)$ are used.
The second term is
\begin{eqnarray}
&&\frac{1}{16\pi}\int\int d\omega_6d\omega_5\hat{a}_s(\omega_6)\hat{a}_i(\omega_5)e^{-i\omega_6 t_6}e^{-i\omega_5 t_5}(e^{-i\omega_6 (\tau_1+\tau_2)}+e^{-i\omega_6 \tau_2}-e^{-i\omega_6 \tau_1}+1)\nonumber\\
&&\times (e^{-i\omega_5 (\tau_1+\tau_2)}-e^{-i\omega_5 \tau_2}+e^{-i\omega_5 \tau_1}+1)\int \int d\omega_s d\omega_i f(\omega_s, \omega_i)\hat{a}_s^\dag(\omega_s)\hat{a}_i^\dag(\omega_i)|0\rangle \nonumber\\
=&& \frac{1}{16\pi} \int \int d\omega_6 d\omega_5 e^{-i\omega_6 t_6}e^{-i\omega_5 t_5} f(\omega_6, \omega_5)(e^{-i\omega_6 (\tau_1+\tau_2)}+e^{-i\omega_6 \tau_2}-e^{-i\omega_6 \tau_1}+1)
\nonumber\\
&&\times (e^{-i\omega_5 (\tau_1+\tau_2)}-e^{-i\omega_5 \tau_2}+e^{-i\omega_5 \tau_1}+1)|0\rangle \nonumber\\
\end{eqnarray}
Combine these two terms:
\begin{eqnarray}
\hat{E}_6^{(+)}\hat{E}_5^{(+)}|\Psi \rangle=&&\frac{1}{16\pi}\int \int d\omega_1d\omega_2e^{-i\omega_5 t_5}e^{-i\omega_6 t_6}[f(\omega_5, \omega_6)(e^{-i\omega_5 (\tau_1+\tau_2)}+e^{-i\omega_5 \tau_1}+e^{-i\omega_5 \tau_2}-1)\nonumber\\
&& \times (e^{-i\omega_6 (\tau_1+\tau_2)}-e^{-i\omega_6 \tau_1}-e^{-i\omega_6 \tau_2}-1)+f(\omega_6, \omega_5)(e^{-i\omega_6 (\tau_1+\tau_2)}+e^{-i\omega_6 \tau_2}-e^{-i\omega_6 \tau_1}+1)\nonumber\\
&& \times(e^{-i\omega_5 (\tau_1+\tau_2)}-e^{-i\omega_5 \tau_2}+e^{-i\omega_5 \tau_1}+1)]|0\rangle
\end{eqnarray}
Then,
\begin{eqnarray}
\langle \Psi |\hat{E}_5^{(-)}\hat{E}_6^{(-)} && \hat{E}_6^{(+)}\hat{E}_5^{(+)}|\Psi \rangle=\left(\frac{1}{16\pi}\right)^2 \int \int d\omega_5^{'}d\omega_6^{'}e^{-i\omega_5^{'} t_5}e^{-i\omega_6^{'} t_6}\nonumber\\
&&\times[f^{*}(\omega_6^{'}, \omega_5^{'})(e^{i\omega_6^{'} (\tau_1+\tau_2)}+e^{i\omega_6^{'} \tau_2}-e^{i\omega_6^{'} \tau_1}+1) (e^{i\omega_5^{'} (\tau_1+\tau_2)}-e^{i\omega_5^{'} \tau_2}+e^{i\omega_5^{'} \tau_1}+1)\nonumber\\
&&+f^{*}(\omega_5^{'}, \omega_6^{'})(e^{i\omega_5^{'} (\tau_1+\tau_2)}+e^{i\omega_5^{'} \tau_1}+e^{i\omega_5^{'} \tau_2}-1)(e^{i\omega_6^{'} (\tau_1+\tau_2)}-e^{i\omega_6^{'} \tau_1}-e^{i\omega_6^{'} \tau_2}-1)\nonumber\\
&&\times \left(\frac{1}{16\pi}\right)^2 \int \int d\omega_5 d\omega_6 e^{-i\omega_5 t_5}e^{-i\omega_6 t_6}\nonumber\\
&&\times[f(\omega_6, \omega_5)(e^{-i\omega_6 (\tau_1+\tau_2)}+e^{-i\omega_6 \tau_2}-e^{-i\omega_6 \tau_1}+1) (e^{-i\omega_5 (\tau_1+\tau_2)}-e^{-i\omega_5 \tau_2}+e^{-i\omega_5 \tau_1}+1)\nonumber\\
&&+f(\omega_5, \omega_6)(e^{-i\omega_5 (\tau_1+\tau_2)}+e^{-i\omega_5 \tau_1}+e^{-i\omega_5 \tau_2}-1)(e^{-i\omega_6 (\tau_1+\tau_2)}-e^{-i\omega_6 \tau_1}-e^{-i\omega_6 \tau_2}-1)]
\end{eqnarray}
Finally,
\begin{eqnarray}
\label{R12}
R(\tau_1,\tau_2)= \int \int d t_5 d t_6 \langle \Psi |\hat{E}_5^{(-)}\hat{E}_6^{(-)} \hat{E}_6^{(+)}\hat{E}_5^{(+)}|\Psi \rangle=\left(\frac{1}{16\pi}\right)^2\int \int d\omega_5d\omega_6 d\omega_5^{'}d\omega_6^{'}\delta(\omega_5-\omega_5^{'})\delta(\omega_6-\omega_6^{'})\nonumber\\
\times[f^{*}(\omega_6^{'}, \omega_5^{'})(e^{i\omega_6^{'} (\tau_1+\tau_2)}+e^{i\omega_6^{'} \tau_2}-e^{i\omega_6^{'} \tau_1}+1) (e^{i\omega_5^{'} (\tau_1+\tau_2)}-e^{i\omega_5^{'} \tau_2}+e^{i\omega_5^{'} \tau_1}+1)\nonumber\\
+f^{*}(\omega_5^{'}, \omega_6^{'})(e^{i\omega_5^{'} (\tau_1+\tau_2)}+e^{i\omega_5^{'} \tau_1}+e^{i\omega_5^{'} \tau_2}-1)(e^{i\omega_6^{'} (\tau_1+\tau_2)}-e^{i\omega_6^{'} \tau_1}-e^{i\omega_6^{'} \tau_2}-1)
\nonumber\\
\times[f(\omega_6, \omega_5)(e^{-i\omega_6 (\tau_1+\tau_2)}+e^{-i\omega_6 \tau_2}-e^{-i\omega_6 \tau_1}+1) (e^{-i\omega_5 (\tau_1+\tau_2)}-e^{-i\omega_5 \tau_2}+e^{-i\omega_5 \tau_1}+1)\nonumber\\
+f(\omega_5, \omega_6)(e^{-i\omega_5 (\tau_1+\tau_2)}+e^{-i\omega_5 \tau_1}+e^{-i\omega_5 \tau_2}-1)(e^{-i\omega_6 (\tau_1+\tau_2)}-e^{-i\omega_6 \tau_2}-e^{-i\omega_6 \tau_1}-1)]\nonumber\\
=\frac{1}{64}\int \int d\omega_5d\omega_6 \times|f(\omega_6, \omega_5)(e^{-i\omega_6 (\tau_1+\tau_2)}+e^{-i\omega_6 \tau_2}-e^{-i\omega_6 \tau_1}+1) (e^{-i\omega_5 (\tau_1+\tau_2)}-e^{-i\omega_5 \tau_2}+e^{-i\omega_5 \tau_1}+1)\nonumber\\
+f(\omega_5, \omega_6)(e^{-i\omega_5 (\tau_1+\tau_2)}+e^{-i\omega_5 \tau_1}+e^{-i\omega_5 \tau_2}-1)(e^{-i\omega_6 (\tau_1+\tau_2)}-e^{-i\omega_6 \tau_1}-e^{-i\omega_6 \tau_2}-1)|^2.
\end{eqnarray}
In above calculation, the relationship of $\delta(\omega-\omega_{'})=\frac{1}{2\pi}\int_{-\infty}^{\infty}e^{i(\omega-\omega^{'})t}dt$ is used. $f^*$ is the complex conjugate of $f$.
In order to introduce less variables, Eq.(\ref{R12}) can be rewritten as
\begin{equation}
R(\tau_1,\tau_2)= \frac{1}{64}\int_{0}^{\infty}\int_{0}^{\infty}d\omega_s d\omega_i r(\omega_s,\omega_i,\tau_1,\tau_2).
\end{equation}
This is Eq.(\ref{R}) in the main context.

\section{The coincidence count rates for the combination interferometer derived from the time domain}
In order to obtain the temporal expression of coincidence count rates, we need to take the Fourier transform of Eq.(\ref{R12}) back into the time representation, yielding to the JTI
\begin{eqnarray}
\label{FT-r}
|\Psi(t_s,t_i)|^2= &&|A_1(t_s, t_i)-A_2(t_s+\tau_1, t_i+\tau_1)+A_3(t_s, t_i+\tau_2)
-A_4(t_s+\tau_2, t_i)+A_5(t_s+\tau_1, t_i+(\tau_1+\tau_2))\nonumber\\
 &&-A_6(t_s+(\tau_1+\tau_2), t_i+\tau_1)+A_7(t_s+(\tau_1+\tau_2), t_i+(\tau_1+\tau_2))-A_8(t_s+\tau_2, t_i+\tau_2)|^2.
\end{eqnarray}
where $\Psi$ is called as effective two-photon wave function. $A$ denotes the quantum-mechanical probability amplitude for the coincidence detection event and can be obtained from the Fourier transform of the JSA at the coordinates marked in Eq.(\ref{FT-r}). After taking module square of Eq.(\ref{FT-r}), the JTI has 64 terms,
\begin{eqnarray}
\label{FT-64}
&&|\Psi(t_s,t_i)|^2=|A_1|^2+|A_2|^2+|A_3|^2+|A_4|^2+|A_5|^2+|A_6|^2+|A_7|^2+|A_8|^2\nonumber\\
&&+A_1^*A_3+A_1A_3^*+A_1^*A_5+A_1A_5^*+A_1^*A_7+A_1A_7^*
+A_2^*A_4+A_2A_4^*+A_2^*A_6+A_2A_6^*+A_2^*A_8+A_2A_8^*\nonumber\\
&&+A_3^*A_5+A_3A_5^*+A_3^*A_7+A_3A_7^*+A_4^*A_6+A_4A_6^*+A_4^*A_8+A_4A_8^*
+A_5^*A_7+A_5A_7^*+A_6^*A_8+A_6A_8^*\nonumber\\
&&-A_1^*A_2-A_1A_2^*-A_1^*A_4-A_1A_4^*-A_1^*A_6-A_1A_6^*-A_1^*A_8-A_1A_8^*-A_2^*A_3-A_2A_3^*-A_2^*A_5-A_2A_5^*\nonumber\\
&&-A_2^*A_7-A_2A_7^*-A_3^*A_4-A_3A_4^*-A_3^*A_6-A_3A_6^*-A_3^*A_8-A_3A_8^*-A_4^*A_5-A_4A_5^*-A_4^*A_7-A_4A_7^*\nonumber\\
&&-A_5^*A_6-A_5A_6^*-A_5^*A_8-A_5A_8^*-A_6^*A_7-A_6A_7^*-A_7^*A_8-A_7A_8^*.
\end{eqnarray}
Then, the coincidence count rates between two detectors as functions of delay time $\tau_1,\tau_2$ can be rewritten as
\begin{equation}
\label{FT-R}
R(\tau_1,\tau_2)=\int_{-\infty}^{\infty}\int_{-\infty}^{\infty}dt_s dt_i |\Psi(t_s, t_i)|^2
\end{equation}
As an example, we take the JSA as the product of two Gaussian functions,
\begin{equation}
\label{f}
f(\omega_s, \omega_i)=\exp\left(- \frac{(\omega_s+\omega_i-\omega_p)^2}{4\sigma_{+}^2}\right)\exp\left(- \frac{(\omega_s-\omega_i)^2}{4\sigma_{-}^2}\right)
\end{equation}
Taking the Fourier transform of Eq.(\ref {f}) at the coordinates marked in Eq.(\ref {FT-r}) and integrating the first eight terms in Eq.(\ref {FT-64}) over the entire time range, we can obtain a normalized factor $8\pi\sigma_{+}\sigma_{-}$. Similarly, integrating $-A_3^*A_6-A_3A_6^*-A_4^*A_5-A_4A_5^*$, $-A_1^*A_8-A_1A_8^*-A_2^*A_7-A_2A_7^*$, $-A_3^*A_4-A_3A_4^*-A_5^*A_6-A_5A_6^*$, and $A_1^*A_7+A_1A_7^*+A_2^*A_8+A_2A_8^*$ in Eq.(\ref {FT-64}), we can obtain the normalized results, which correspond to the second, the third, the fourth and the last two terms in Eq.(\ref {Rn}), respectively. The remaining terms in Eq.(\ref {FT-64}) have no contributions to the integral of Eq.(\ref {FT-R}). We therefore obtain the same result from the time domain as Eq.(\ref {Rn}) derived from the frequency domain.

\section{The coincidence count rates for the HOMI and the NOONI}
In order to compare the results of the HOMI and the NOONI with the combination interferometer, we give the coincidence count rates of these two interferometers as follows (see details in \cite{JIN2018})
\begin{equation}
\label{R+-}
R_{\pm}(\tau)\sim \int_{0}^{\infty}\int_{0}^{\infty}d\omega_s d\omega_i r_{\pm}(\omega_s,\omega_i,\tau).
\end{equation}
where
\begin{equation}
\label{r+}
r_{+}(\omega_s,\omega_i,\tau)=|f(\omega_i,\omega_s)(e^{-i\omega_i \tau}+1)(e^{-i\omega_s \tau}+1)+f(\omega_s,\omega_i)(e^{-i\omega_s \tau}-1)(e^{-i\omega_i \tau}-1)|^2,
\end{equation}
\begin{equation}
\label{r-}
r_{-}(\omega_s,\omega_i,\tau)=|f(\omega_i,\omega_s)e^{-i\omega_s \tau}-f(\omega_s,\omega_i)e^{-i\omega_i \tau}|^2.
\end{equation}
The subscripts $``+"$ and $``-"$ corespond to a NOON state and standard HOM interferometer, respectively. If the JSA is symmetric, i.e., $f(\omega_s, \omega_i)=f(\omega_i, \omega_s)$ , we can further rewrite the above equation as
\begin{equation}
\label{r+-}
r_{\pm}(\omega_s,\omega_i,\tau)=|f(\omega_s,\omega_i)|^2[1\pm \cos(\omega_s \pm \omega_i)\tau].
\end{equation}
If the JSA is antisymmetric, i.e., $f(\omega_s, \omega_i)=-f(\omega_i, \omega_s)$, we have
\begin{equation}
r_{+}(\omega_s,\omega_i,\tau)=r_{-}(\omega_s,\omega_i,\tau)=|f(\omega_s,\omega_i)|^2[1+\cos(\omega_s - \omega_i)\tau].
\end{equation}


\begin{thebibliography}{xyz00}

\bibitem{Pan}
Jian-Wei Pan, Zeng-Bing Chen, Chao-Yang Lu, Harald Weinfurter, Anton Zeilinger, and Marek Żukowski, “Multiphoton entanglement and interferometry,” Rev. Mod. Phys. {\bf 84}, 777 (2012).

\bibitem{PRL1987}
C. K. Hong, Z. Y. Ou, and L. Mandel, ``Measurement of subpicosecond time intervals between two photons by interference,'' Phys. Rev. Lett. {\bf 59}, 2044-2046 (1987).


\bibitem{RPP2021}
Frédéric Bouchard, Alicia Sit, Yingwen Zhang, Robert Fickler, Filippo M Miatto, Yuan Yao, Fabio Sciarrino and Ebrahim Karimi, ``Two-photon interference: the Hong–Ou–Mandel effect,'' Rep. Prog. Phys. {\bf 84}, 012402 (2021).

\bibitem{NOON}
A. N. Boto, P. Kok, D. S. Abrams, S. L. Braunstein, C. P. Williams, and J. P. Dowling, ``Quantum Interferometric Optical Lithography: Exploiting Entanglement to Beat the Diffraction Limit," Phys. Rev. Lett. {\bf 85}, 2733 (2000).

\bibitem{PRL1988}
Z. Y. Ou and L. Mandel, ``Violation of Bell’s inequality and classical probability in a two-photon correlation experiment," Phys. Rev. Lett. {\bf 61}, 50–53 (1988).

\bibitem{PRA1992}
A. M. Steinberg, P. G. Kwiat, and R. Y. Chiao, ``Dispersion cancellation and high-resolution time measurements in a fourth order optical interferometer," Phys. Rev. A {\bf 45}, 6659–6665 (1992).

\bibitem{PRL2009}
O. Minaeva, C. Bonato, B. E. A. Saleh, D. S. Simon, and A. V. Sergienko, ``Odd- and even-order dispersion cancellation in quantum interferometry," Phys. Rev. Lett. {\bf 102}, 100504 (2009).

\bibitem{OE2017}
Jinsoo Ryu, Kiyoung Cho, Cha-Hwan Oh, and Hoonsoo Kang, ``All-order dispersion cancellation and energy-time entangled state,'' Opt. Express {\bf 25}, 1360-1380 (2017).

\bibitem{communication}
N. Gisin and R. Thew, ``Quantum communication," Nat. Photonics {\bf 1}, 165–171 (2007).

\bibitem{computer}
T. D. Ladd, F. Jelezko, R. Laflamme, Y. Nakamura, C. Monroe, and J. L. O’Brien, ``Quantum computers," Nature {\bf 464}, 45–53 (2010).

\bibitem{Imaging1}
Shih, Y. H.,“Quantum Imaging,”  2007, IEEE J. Sel. Top. Quantum
Electronics, {\bf 13}, 1016.

\bibitem{Imaging}
P.-A. Moreau, E. Toninelli, T. Gregory, and M. J. Padgett, ``Imaging with quantum states of light," Nat. Rev. Phys. {\bf 1}, 367–380 (2019).

\bibitem{metrology1}
B. Bell, S. Kannan, A. McMillan, A. S. Clark, W. J. Wadsworth, and J. G. Rarity, ``Multicolor quantum metrology with entangled photons," Phys. Rev. Lett. {\bf 111}, 093603 (2013).

\bibitem{metrology2}
Chen Y, Hong L and Chen L, ``Quantum interferometric metrology
with entangled photons," Front. Phys. {\bf 10}, 892519 (2022).




\bibitem{attosecond}
A. Lyons, G. C. Knee, E. Bolduc, T. Roger, J. Leach, E. M. Gauger, and D. Faccio,``Attosecond-resolution Hong-Ou-Mandel interferometry," Sci. Adv. {\bf 4}, eaap9416 (2018).

\bibitem{NOON1}
K. Edamatsu, R. Shimizu, and T. Itoh, ``Measurement of the Photonic de Broglie Wavelength of Entangled Photon Pairs Generated by Spontaneous Parametric Down-Conversion," Phys. Rev. Lett. {\bf 89}, 213601 (2002)

\bibitem{NOON2}
M. W. Mitchell, J. S. Lundeen, and A. M. Steinberg, ``Super-resolving phase measurements with a multiphoton entangled state," Nature {\bf 429}, 161–164 (2004)

\bibitem{NOON3}
V. Giovannetti, S. Lloyd, and L. Maccone, ``Quantum-enhanced measurements: beating the standard quantum limit," Science {\bf 306}, 1330–1336 (2004).

\bibitem{NOON4}
Nagata, T., R. Okamoto, J. O‘Brien, K. Sasaki, and S. Takeuchi, ``Beating the Standard Quantum Limit with Four-Entangled Photons," Science {\bf 317}, 726 (2007).


\bibitem{microscopy1}
T. Ono, R. Okamoto, and S. Takeuchi, ``An entanglement-enhanced microscope," Nat. Commun. {\bf 4}, 2426 (2013).

\bibitem{microscopy2}
Y. Israel, S. Rosen, and Y. Silberberg, ``Supersensitive polarization microscopy using NOON states of light," Phys. Rev. Lett. {\bf 112}, 103604(2014).

\bibitem{microscopy3}
R.B. Jin, M. Fujiwara, R. Shimizu, R. J. Collins, G. S. Buller, T. Yamashita, S. Miki, H. Terai, M. Takeoka, and M. Sasaki, “Detection dependent six-photon Holland-Burnett state interference,” Sci. Rep. {\bf 6},36914 (2016).

\bibitem{error}
M. Bergmann and P. van Loock, ``Quantum error correction against photon loss using noon states," Phys. Rev. A {\bf 94}, 012311 (2016).



\bibitem{JIN2018}
Rui-Bo Jin, and Ryosuke Shimizu,``Extended Wiener–Khinchin theorem for quantum spectral analysis," Optica {\bf 5}, 93-98 (2018).

\bibitem{PRApplied2022}
Chen Y, Chen L, ``Quantum Wiener-Khinchin Theorem for Spectral-Domain Optical Coherence Tomography," Phys. Rev. Applied {\bf 18}, 014077 (2022).

\bibitem{PRA2002}
Vittorio Giovannetti, Lorenzo Maccone, Jeffrey H. Shapiro, and Franco N. C. Wong, ``Extended phase-matching conditions for improved entanglement generation," Phys. Rev. A {\bf 66}, 043813 (2002).

\bibitem{PRA2013}
Ayman F. Abouraddy, Timothy M. Yarnall, and Giovanni Di Giuseppe, ``Phase-unlocked Hong-Ou-Mandel interferometry," Phys. Rev. A {\bf 87}, 062106 (2013).

\bibitem{PRL2018}
J.-P. W. MacLean, J. M. Donohue, and K. J. Resch, “Direct characterization of ultrafast energy-time entangled photon pairs,” Phys. Rev. Lett.
{\bf 120}, 053601 (2018).

\bibitem{PRApplied2018}
Rui-Bo Jin, Takuma Saito, and Ryosuke Shimizu, ``Time-Frequency Duality of Biphotons for Quantum Optical Synthesis," Phys. Rev. Applied {\bf 10}, 034011 (2018).

\bibitem{JMO2018}
Kevin Zielnicki, Karina Garay-Palmett, Daniel Cruz-Delgado, Hector Cruz-Ramirez, Michael F. O’Boyle, Bin Fang, Virginia O. Lorenz, Alfred B. U’Ren and Paul G. Kwiat, ``Joint spectral characterization of photon-pair sources, Journal of Modern Optics," 65:10, 1141-1160 (2018)

\bibitem{PRApplied2021}
Y. Mukai ,M. Arahata ,T. Tashima, R. Okamoto, and S. Takeuchi, ``Quantum Fourier-Transform Infrared Spectroscopy for Complex Transmittance Measurements," Phys. Rev. Applied {\bf 15}, 034019 (2021).

\bibitem{PRA2019}
Jean-Philippe W. MacLean, Sacha Schwarz, and Kevin J. Resch, ``Reconstructing ultrafast energy-time-entangled two-photon pulses," Phys. Rev. A {\bf 100}, 033834(2019).

\bibitem{Optica2020}
Alex O. C. Davis, Valérian Thiel, and Brian J. Smith, "Measuring the quantum state of a photon pair entangled in frequency and time," Optica 7, 1317-1322 (2020)


\bibitem{Walborn03}
S. P. Walborn, A. N. de Oliveira, S. Pádua, and C. H. Monken, ``Multimode Hong-Ou-Mandel Interference," Phys. Rev. Lett. 90, 143601 (2003)


\bibitem{Optica20}
S. Francesconi, F. Baboux, A. Raymond, N. Fabre, G. Boucher, A. Lemaître, P. Milman, M. Amanti, and S. Ducci, ``Engineering two-photon wavefunction and exchange statistics in a semiconductor chip," Optica 7, 316 (2020).

\bibitem{Wang20}
Liu, Zhi-Feng and Chen, Chao and Xu, Jia-Min and Cheng, Zi-Mo and Ren, Zhi-Cheng and Dong, Bo-Wen and Lou, Yan-Chao and Yang, Yu-Xiang and Xue, Shu-Tian and Liu, Zhi-Hong and Zhu, Wen-Zheng and Wang, Xi-Lin and Wang, Hui-Tian, ``Hong-Ou-Mandel Interference between Two Hyperentangled Photons Enables Observation of Symmetric and Antisymmetric Particle Exchange Phases," Phys. Rev. Lett. 129, 263602 (2022)

\bibitem{Fabre22}
Fabre, N., ``Interferometric signature of different spectral symmetries of biphoton states," Phys. Rev. A 105, 053716 (2022)

\bibitem{SR-PRA2015}
T. Gerrits, F. Marsili, V. B. Verma, L. K. Shalm, M. Shaw, R. P. Mirin,and S. W. Nam, ``Spectral correlation measurements at the Hong-Ou-Mandel interference dip,'' Phys. Rev. A {\bf 91}, 013830 (2015).

\bibitem{SR-OE2015}
R.-B. Jin, T. Gerrits, M. Fujiwara, R. Wakabayashi, T. Yamashita, S.
Miki, H. Terai, R. Shimizu, M. Takeoka, and M. Sasaki, ``Spectrally
resolved Hong-Ou-Mandel interference between independent photon
sources,'' Opt. Express {\bf 23}, 28836 (2015).

\bibitem{jin2016}
R.-B. Jin, R. Shimizu, M. Fujiwara, M. Takeoka, R. Wakabayashi, T. Yamashita, S. Miki, H. Terai, T. Gerrits, and M. Sasaki, ``Simple method of generating
and distributing frequency-entangled qudits," Quantum Sci. Technol. {\bf 1}, 015004 (2016).

\bibitem{npj2021}
Yuanyuan Chen, Sebastian Ecker, Lixiang Chen, Fabian Steinlechner, Marcus Huber, Rupert Ursin, ``Temporal distinguishability in Hong-Ou-Mandel interference: Generation and characterization of high-dimensional frequency entanglement,'' Npj.quantum inform {\bf 7}, 167(2021).

\bibitem{OLT2023}
Baihong Li, Boxin Yuan, Changhua Chen, Xiao Xiang, Runai Quan, Ruifang Dong, Shougang Zhang, Rui-Bo Jin, ``Spectrally resolved two-photon interference in a modified Hong–Ou–Mandel interferometer,'' Opt Laser Technol {\bf 159}, 109039(2023).


\bibitem{3-photons2019}
Venkata Vikram Orre, Elizabeth A. Goldschmidt, Abhinav Deshpande, Alexey V. Gorshkov, Vincenzo Tamma, Mohammad Hafezi, and Sunil Mittal,``Interference of Temporally Distinguishable Photons Using Frequency-Resolved Detection'' Phys. Rev. Lett. {\bf 123}, 123603 (2019).

\bibitem{SR-NOON2021}
Rui-Bo Jin, Ryosuke Shimizu, Takafumi Ono, Mikio Fujiwara, Guang-Wei Deng, Qiang Zhou, Masahide Sasaki, Masahiro Takeoka, ``Spectrally resolved NOON state interference," arXiv: 2104.01062

\end{thebibliography}
\end{document}